\documentclass[aps,prd,preprint,floatfix,superscriptaddress,nofootinbib,longbibliography,a4paper]{revtex4-1}
\pdfoutput=1
\usepackage{color}
\usepackage{graphicx}
\usepackage{textcomp}
\usepackage{subfigure}
\usepackage{feynmp}
\usepackage{amsmath,amssymb,array}
\usepackage{enumerate}
\usepackage{slashed}
\usepackage{hhline}
\usepackage{hyperref}
\hypersetup{colorlinks=true,
	linkcolor=blue,
	filecolor=magenta,      
	urlcolor=blue}
\usepackage{tcolorbox}
\definecolor{bred}{rgb}{1.0,0.0,0.0}
\definecolor{bgreen}{rgb}{0.0,1.0,0.0}
\newtcbox{\rb}{on line,
  colframe=bred,colback=bred!15!white,
  boxrule=0.0pt,arc=3pt,boxsep=0pt,left=2pt,right=2pt,top=2pt,bottom=2pt}
\newtcbox{\gb}{on line,
  colframe=bgreen,colback=bgreen!15!white,
  boxrule=0.0pt,arc=3pt,boxsep=0pt,left=2pt,right=2pt,top=2pt,bottom=2pt}

\newcommand{\beqa}{\begin{eqnarray}}
\newcommand{\eeqa}{\end{eqnarray}}
\newcommand{\be}{\begin{equation}}
\newcommand{\ee}{\end{equation}}
\newcommand{\ba}{\begin{array}} 
\newcommand{\ea}{\end{array}}

\begin{document} 
\vspace*{0.5cm}
\title{Radiatively generated fermion mass hierarchy from flavour non-universal gauge symmetries}
\bigskip
\author{Gurucharan Mohanta}
\email{gurucharan@prl.res.in}
\affiliation{Theoretical Physics Division, Physical Research Laboratory, Navarangpura, Ahmedabad-380009, India}
\affiliation{Indian Institute of Technology Gandhinagar, Palaj-382355, India\vspace*{0.5cm}}
\author{Ketan M. Patel}
\email{kmpatel@prl.res.in}
\affiliation{Theoretical Physics Division, Physical Research Laboratory, Navarangpura, Ahmedabad-380009, India}

\begin{abstract}
A framework based on a class of abelian gauge symmetries is proposed in which the masses of only the third generation quarks and leptons arise at the tree level. The fermions of the first and second families receive their masses through radiative corrections induced by the new gauge bosons in the loops. It is shown that the class of abelian symmetries which can viably implement this mechanism are flavour non-universal in nature. Taking the all-fermion generalization of the well-known leptonic $L_\mu-L_\tau$ and $L_e - L_\mu$ symmetries, we construct an explicit renormalizable model based on two $U(1)$ which is shown to reproduce the observed fermion mass spectrum of the Standard Model. The first and second generation fermion masses are loop suppressed while the hierarchy between these two generations results from a gap between the masses of two vector bosons of the extended gauge symmetries. Several phenomenological aspects of the flavourful new physics are discussed and lower limits on the masses of the vector bosons are derived.
\end{abstract}

\maketitle

\section{Introduction}
\label{sec:intro}
Masses of the elementary fermions in the Standard Model (SM) are incalculable parameters of the theory and their values are determined entirely from the observations only. This leaves several unanswered questions about the observed pattern of the masses and mixings of quarks and leptons \cite{Feruglio:2015jfa}. One of the elegant ways to understand the peculiar hierarchical masses of the charged fermions is to let only the third generation fermions become massive at the leading order while the masses of the first two generations arise through quantum corrections. In this way, the masses of the first and second generation fermions may be made completely or  partially calculable parameters of the theory. Such an approach was considered in \cite{Barr:1978rv,Wilczek:1978xi,Yanagida:1979gs} soon after the inception of the SM. Subsequent attempts in this direction include generating radiative masses through loops involving new scalars and fermions \cite{Balakrishna:1987qd,Balakrishna:1988ks,Balakrishna:1988xg,Balakrishna:1988bn,Babu:1988fn,Babu:1989tv,Graham:2009gr,Chiang:2021pma,Chiang:2022axu,Baker:2020vkh,Baker:2021yli}, new scalars and top-quark \cite{Dobrescu:2008sz} and new gauge bosons in the extended gauge theories \cite{Weinberg:2020zba,Jana:2021tlx}.

In this paper, we explore a framework in a similar direction but based on a different class of gauge symmetry leading to very different results both qualitatively and quantitatively. The extended gauge symmetry is abelian and it ensures that only the third generation quarks and leptons get masses at the leading order. Subsequently, the masses of the first and second generation fermions are induced by radiative corrections through the vector bosons of extended symmetry and heavy massive fermions in the loop and, therefore, are suppressed. The masses of the first two generation fermions become calculable parameters albeit they are expressed in terms of some undetermined parameters of the theory which are not required to take arbitrarily small values. Through general analysis, we show that the new abelian gauge symmetry must be flavour non-universal to the basic mechanism to work viably. In turn, this leads to flavourful new physics in both the quark and lepton sectors offering phenomenologically rich possibilities. We construct an explicit model based on two $U(1)$s which are generalizations of the well-known leptonic $L_\mu - L_\tau$ and $L_e - L_\mu$ symmetries and are applied to all the fermions of a given generation. It is shown through detailed numerical studies that the model is capable of reproducing the realistic fermion mass spectrum and it overcomes many inviable predictions found in a recent proposal along a similar direction but based on a different gauge symmetry \cite{Weinberg:2020zba}.

The rest of the paper is organized as follows. The basic mechanism is discussed in section \ref{sec:general}. An explicit model based on it is outlined in section \ref{sec:model}. In section \ref{sec:solutions}, we give example solutions which reproduce the observed fermion mass spectrum and discuss the various phenomenological implications in section \ref{sec:pheno}. The study is summarized in section \ref{sec:summary}.

\section{General Framework}
\label{sec:general}
We first discuss the basic mechanism leading to the massive third generation at the tree level and radiative masses for the lighter generations. Consider a toy framework with three generations of chiral fermions, $f^\prime_{L i}$ and $f^\prime_{R i}$ ($i=1,2,3$), and a pair of vectorlike fermions $F^\prime_{L,R}$. The fermions are charged under a $U(1)$ gauge symmetry with the following interaction term:
\be \label{L_gauge}
-{\cal L}_{\rm gauge} = g_X X_\mu \left(q_{L \alpha}\, \overline{f}^\prime_{L \alpha} \gamma^\mu f^\prime_{L \alpha} + q_{R \alpha}\, \overline{f}^\prime_{R \alpha} \gamma^\mu f^\prime_{R \alpha} \right)\,, \ee
where $\alpha=1,...,4$, $f^\prime_{L \alpha} = (f^\prime_{L i},F^\prime_L)$, $f^\prime_{R\alpha} = (f^\prime_{R i},F^\prime_R)$ and $q_{L4} = q_{R4}$.

The mass Lagrangian in this basis is written as
\be \label{L_mass}
-{\cal L}_{m} = \overline{f}^\prime_{L \alpha}\, {\cal M}^{(0)}_{\alpha \beta}\, f^\prime_{R \beta} + {\rm h.c.}\,, \ee
where the $4 \times 4$ tree-level mass matrix ${\cal M}^{(0)}$ has a particular form
\be \label{M}
{\cal M}^{(0)} = \left( \ba{cc} 0_{3 \times 3} & (\mu)_{3 \times 1} \\ (\mu^\prime)_{1 \times 3} & m_F \ea \right)\,.\ee 
Here, $\mu = (\mu_1,\mu_2,\mu_3)^T$ and $\mu^\prime = (\mu^\prime_1, \mu^\prime_2, \mu^\prime_3)$. The above form of ${\cal M}$ results into a pair of massive and two massless fermions. For heavy vectorlike fermions, i.e. $m_F \gg \mu_i, \mu^\prime_i$, the effective $3 \times 3$ mass matrix can be obtained as
\be \label{M_eff}
M^{(0)}_{ij} = -\frac{1}{m_F} \mu_i\,\mu^\prime_j \,. \ee
The above matrix is of rank one and has two vanishing eigenvalues. In this way, only the third generation is arranged to acquire a tree level mass in the underlying framework.

To compute higher order corrections to tree-level masses, we obtain the physical basis using
\be \label{fprime}
f^\prime_{L,R} = {\cal U}_{L,R}\,f_{L, R}\,, \ee
where the unprimed fields are in physical basis and ${\cal U}_{L,R}$ are $4 \times 4$ unitary matrices which can be obtained using
\be \label{M_diag}
{\cal U}_L^\dagger\,{\cal M}^{(0)}\,{\cal U}_R = {\cal D} \equiv {\rm Diag.}(0,0,m_3,m_4)\,.\ee
The gauge interactions in Eq. (\ref{L_gauge}), in the new basis, becomes
\be \label{L_gauge_2}
-{\cal L}_{\rm gauge} = g_X X_\mu \left(({\cal Q}_L)_{\alpha \beta}\, \overline{f}_{L \alpha} \gamma^\mu f_{L \beta} + ({\cal Q}_R)_{\alpha \beta}\, \overline{f}_{R \alpha} \gamma^\mu f_{R \beta} \right)\,, \ee
where
\be \label{calQ}
{\cal Q}_{L,R} = {\cal U}_{L,R}^\dagger\,q_{L,R}\,{\cal U}_{L,R}\,,\ee
and $q_{L} = {\rm Diag.}(q_{L 1},q_{L 2},...)$ and so on. The matrices ${\cal Q}_{L,R}$ are not diagonal in general.

At 1-loop order, the fermions of the first and second generations can receive masses through diagrams involving the gauge boson and massive fermions in the loop. The one-particle-irreducible two-point function evaluated in the Feynman-'t Hooft gauge gives the following results.
\be \label{Sigma}
\Sigma_{\alpha \beta}(p=0) = \sigma^L_{\alpha \beta}\, P_L + \sigma^R_{\alpha \beta}\, P_R\,,\ee
with
\beqa \label{sigma_LR}
\sigma^L_{\alpha \beta} & = &   \frac{g_X^2}{4 \pi^2}\, \sum_\gamma ({\cal Q}_R)_{\alpha \gamma}\, ({\cal Q}_L)_{\gamma \beta}\, m_\gamma\, B_0[M_X^2,m_\gamma^2]\,, \nonumber \\
\sigma^R_{\alpha \beta} & = &   \frac{g_X^2}{4 \pi^2}\, \sum_\gamma ({\cal Q}_L)_{\alpha \gamma}\,({\cal Q}_R)_{\gamma \beta}\,m_\gamma \, B_0[M_X^2,m_\gamma^2]\,,\eeqa
where 
\be \label{B0}
B_0[M^2,m^2] \equiv \Delta_\epsilon - \frac{M^2 \ln M^2 - m^2 \ln m^2}{M^2 - m^2}\,,\ee
is Passarino-Veltmann function and 
\be \label{Depsilon}
\Delta_\epsilon = \frac{2}{\epsilon} + 1 - \gamma + \ln 4\pi\,,\ee
is divergent part of loop integration when $\epsilon \to 0$. The explicit derivation of the above result is given in Appendix \ref{app:loop}.

The 1-loop corrected fermion mass matrix obtained using Eqs. (\ref{M_diag},\ref{sigma_LR}) can be written as
\be \label{M_loopcorr}
{\cal M} = {\cal M}^{(0)} + \delta{\cal M}\,,\ee
where
\be \label{del_calM}
\delta{\cal M} = {\cal U}_L\,\sigma^R\,{\cal U}_R^\dagger\,. \ee
In general, the loop contributions $\delta{\cal M}$ has divergent terms proportional to $\Delta_\epsilon$. The renormalizability requires that the $3 \times 3$ upper-left block of $\delta{\cal M}$ should be finite as there are no counterterms that could remove the divergences. Denoting the divergent part of $\delta{\cal M}$ as $\delta{\cal M}_{\rm div}$, we find
\be \label{DM_div}
\delta{\cal M}_{\rm div} \propto {\cal U}_L\, {\cal Q}_L\, {\cal D}\, {\cal Q}_R\, {\cal U}_R^\dagger = q_L\,{\cal M}^{(0)}\,q_R\,, \ee
where the last equality follows from Eqs. (\ref{M_diag},\ref{calQ}). Using ${\cal M}$ given in Eq. (\ref{M}) and since $q_{L,R}$ are diagonal matrices, one finds
\be \label{DM_div2}
\left(\delta{\cal M}_{\rm div}\right)_{ij} = 0\,.\ee
Therefore, the $3 \times 3$ upper-left block of $\delta{\cal M}$ is finite as it is expected from the renormalizabilty \cite{Barr:1978rv,Weinberg:1972ws}.

The finite part of $\delta {\cal M}$ is of our main interest and it can be simplified to
\be \label{del_calM_2}
\left(\delta{\cal M} \right)_{\alpha \beta} = \frac{g_X^2}{4 \pi^2} q_{L \alpha} q_{R \beta}\, \sum_\gamma \left({\cal U}_L\right)_{\alpha \gamma} \left({\cal U}_R^*\right)_{\beta \gamma}\, m_\gamma\,b_0[M_X^2,m_\gamma^2]\,, \ee 
where 
\be \label{b0}
b_0[M^2,m^2] \equiv -\frac{M^2 \ln M^2 - m^2 \ln m^2}{M^2 - m^2}\,.\ee
Further simplification is possible in the seesaw approximation, $m_F \gg \mu_i, \mu^\prime_i$. In this case, ${\cal U}_{L,R}$ can be written in the following form \cite{Joshipura:2019qxz}
\be \label{U_block}
{\cal U}_{L,R} \simeq \left(\ba{cc} U_{L,R} & -\rho_{L,R} \\ \rho_{L,R}^\dagger U_{L,R} & 1\ea\right)\,,\ee
where $\rho_L = -m_F^{-1}\mu$ and $\rho_R^\dagger = -m_F^{-1}\mu^\prime$ are $3 \times 1$ and $1 \times 3$ are matrices, respectively. $U_{L,R}$ are $3 \times 3$ matrices that diagonalize $M^{(0)}$ given in Eq. (\ref{M_eff}). Explicitly,
\be \label{Meff_diag}
U_L^\dagger\,M^{(0)}\,U_R = {\rm Diag.}(0,0,m_3)\,.\ee
Using the definition Eq. (\ref{M_eff}), the above equation can also be written as
\be \label{Meff_diag_comp}
\left(U_L\right)_{i3} \left(U_R^*\right)_{j3}\, m_3\, = M^{(0)}_{ij} = -\frac{1}{m_F}\mu_i \mu^\prime_j\,.\ee

Substituting Eqs. (\ref{U_block},\ref{Meff_diag}) in Eq. (\ref{del_calM_2}), we find
\beqa \label{del_calM_33}
\left(\delta{\cal M} \right)_{ij} &\simeq& \frac{g_X^2}{4 \pi^2} q_{L i} q_{R j}\, \left(U_L\right)_{i3} \left(U_R^*\right)_{j3}\, m_3\,\left(b_0[M_X^2,m_3^2] - b_0[M_X^2,m_F^2] \right)\,,\nonumber \\
& = & \frac{g_X^2}{4 \pi^2} q_{L i} q_{R j}\, M^{(0)}_{ij}\left(b_0[M_X^2,m_3^2] - b_0[M_X^2,m_F^2] \right)\,,\eeqa
with $i,j=1,2,3$ and repeated indices are not summed. The second line in the above equation follows from Eq. (\ref{Meff_diag_comp}). The correction to the remaining elements, after some straight-forward algebraic simplification, are obtained as
\beqa \label{del_calM_44}
\left(\delta{\cal M} \right)_{i4} &\simeq & \frac{g_X^2}{4 \pi^2} q_{L i} q_{R 4}\, \mu_i\, \left(b_0[M_X^2,m_F^2] + \sum_j \frac{|\mu^\prime_j|^2}{m_F^2}\, b_0[M_X^2,m_3^2]\right)\,,\nonumber \\
\left(\delta{\cal M} \right)_{4i} &\simeq & \frac{g_X^2}{4 \pi^2} q_{L 4} q_{R i}\, \mu^\prime_i\, \left(b_0[M_X^2,m_F^2] + \sum_j \frac{|\mu_j|^2}{m_F^2}\, b_0[M_X^2,m_3^2]\right)\,,\nonumber \\
\left(\delta{\cal M} \right)_{44} &\simeq & \frac{g_X^2}{4 \pi^2} q_{L 4} q_{R 4}\,m_F\, \left(b_0[M_X^2,m_F^2] -  \frac{m_3^2}{m_F^2}\, b_0[M_X^2,m_3^2]\right)\,. \eeqa

Using the above results, the 1-loop corrected fermion mass matrix, Eq. (\ref{M_loopcorr}), can be parametrized as
\be \label{M_corr}
{\cal M} = \left( \ba{cc} (\delta M)_{3 \times 3} & (\tilde{\mu})_{3 \times 1} \\ (\tilde{\mu}^\prime)_{1 \times 3} & \tilde{m}_F \ea \right)\,,\ee
with 
\be \label{DeltaM}
\left(\delta M \right)_{ij} = \left(\delta{\cal M} \right)_{ij}\,,~\tilde{\mu}_i = \mu_i + \left(\delta{\cal M} \right)_{i4}\,,~\tilde{\mu}^\prime_i = \mu^\prime_i + \left(\delta{\cal M} \right)_{4i}\,,~\tilde{m}_F = m_F +\left(\delta{\cal M} \right)_{44}\,.\ee
In the seesaw approximation, $\delta M_{ij} \ll \tilde{\mu}_i, \tilde{\mu}^\prime_i \ll \tilde{m}_F$, the effective $3 \times 3$ mass matrix for the lighter fermions can then be written as
\be \label{M_eff_corr}
M = \delta M - \frac{1}{\tilde{m}_F}\tilde{\mu} \tilde{\mu}^{\prime}\,.\ee
Comparing the above with Eq. (\ref{M_eff}), it is noticed that the second term is similar to $M^{(0)}$ with original elements replaced by their 1-loop corrected values. This contribution is still of rank-1 and contributes only to the masses of the third generation. The first term can induce masses for the first and/or second generation fermions depending on the chosen $U(1)$ charges.

It is important to note that the flavour universal $U(1)$ symmetry cannot induce radiative masses for the lighter generations. This can be understood as the following. For $q_{L1}=q_{L2}=q_{L3}$ and $q_{R1}=q_{R2}=q_{R3}$, one finds $\delta M \propto M^{(0)}$ from Eq. (\ref{del_calM_33}) and $\tilde{\mu} \propto \mu$, $\tilde{\mu^\prime} \propto \mu^\prime$, $\tilde{m_F} \propto m_F$ from Eq. (\ref{del_calM_44}). Altogether, this implies $M \propto M^{(0)}$ and hence the 1-loop corrected mass matrix remains of rank one. Therefore, a flavour non-universal $U(1)$ is necessarily required in order to generate the masses for the first and second generation fermions. We find that this observation of ours is in conflict with the results of \cite{Jana:2021tlx} which uses flavour universal $U(1)_{B-L}$ to induce radiative masses for the first generation fermions.

It can be also noted that for a generic choice of $U(1)$ charges, $\delta M$ could lead to masses of a similar magnitude for the first and second generation fermions. To generate only the second generation fermion masses at 1-loop, one can choose $q_{L1}=q_{R1}=0$. As it can be seen from Eq. (\ref{del_calM_33}), this leads to a massless first generation and loop-suppressed mass for the second generation. Masses for the first generation fermions can be induced similarly by introducing another flavoured $U(1)$ under which the first generation fermions are non-trivially charged. The mass hierarchy between the first and second generations can be arranged by choosing hierarchical masses for the gauge bosons of two $U(1)$s. For example, the loop integration factor for $M_X \gg m_F \gg m_3$ can be simplified as
\be \label{b0_limit}
b_0[M_X^2,m_3^2] - b_0[M_X^2,m_F^2] \simeq -\frac{m_F^2}{M_X^2}\,\ln\frac{m_F^2}{M_X^2}\,.\ee
Therefore, two $U(1)$s with $M_{X_2} \gg M_{X_1}$ can lead to hierarchical masses of first and second generations even though both are generated at 1-loop. A suitable choice of $U(1)$ charges which can achieve this is discussed in the next section.

\section{A model}
\label{sec:model}
As discussed in the last section, the proposed mechanism requires at least two $U(1)$ symmetries under which the three generations of the SM quarks and leptons are charged non-universally. Therefore, we choose $G_F=U(1)_1 \times U(1)_2$ as an extended gauge symmetry for the model. In addition to three generations of the SM quarks and leptons, $Q_{L i} \sim (3,2,\frac{1}{6})$, $u_{R i} \sim (3,1,\frac{2}{3})$, $d_{R i} \sim (3,1,-\frac{1}{3})$, $L_{L i} \sim (1,2,-\frac{1}{2})$ and $e_{R i} \sim (1,1,-1)$, we introduce three copies of a pair of the Higgs doublets, i.e. $H_{u i} \sim (1,2,-\frac{1}{2})$, $H_{d i} \sim (1,2,\frac{1}{2})$ and SM singlets $\eta_i \sim (1,1,0)$. We also introduce a pair of vectorlike fermions in each sector, namely $T_{L,R} \sim (3,1,\frac{2}{3})$, $B_{L,R} \sim (3,1,-\frac{1}{3})$ and $E_{L,R} \sim (1,1,-1)$. In the above, the quantities in the brackets denote the transformation properties under $(SU(3)_C, SU(2)_L, U(1)_Y)$.

Under the new $G_F = U(1)_1 \times U(1)_2$ symmetry, all the first, second and third generation fermions and scalars have charges $(0,1)$, $(1,-1)$ and $(-1,0)$, respectively. In this way, $U(1)_1$ can be identified as ``$2-3$ symmetry" and $U(1)_2$ as ``$1-2$ symmetry". They are generalizations of $L_\mu - L_\tau$ and $L_e - L_\mu$ symmetries, respectively, discussed in the literature in the context of leptons \cite{Foot:1990mn,He:1990pn,He:1991qd}. The vectorlike fermions are neutral under the $G_F$. The fermion and scalar fields and their charges under the SM and $G_F$ are summarized in Table \ref{tab:fields}. It is straightforward to check that the $G_F$ is non-anomalous since a pair of families of fermions and scalars have equal and opposite charges under each $U(1)$. 
\begin{table}[t]
\begin{center}
\begin{tabular}{cccc} 
\hline
\hline
~~Fields~~&~~$(SU(3)_c \times SU(2)_L \times U(1)_Y)$~~&~~$U(1)_1$~~&~~$U(1)_2$~~\\
\hline
$Q_{L_i} $ & $(3,2,\frac{1}{6}) $ & \{0,1,-1\} & \{1,-1,0\}\\
$u_{R_i} $ & $(3,1,\frac{2}{3}) $ & \{0,1,-1\} & \{1,-1,0\}\\
$d_{R_i} $ & $(3,1,-\frac{1}{3}) $ & \{0,1,-1\} & \{1,-1,0\}\\
\hline
$L_{L_i} $ & $(1,2,-\frac{1}{2}) $ & \{0,1,-1\} & \{1,-1,0\}\\
$e_{R_i} $ & $(1,1,-1) $ & \{0,1,-1\} & \{1,-1,0\}\\
\hline
$H_{u_i}$ & $(1,2,-\frac{1}{2}) $ & \{0,1,-1\} & \{1,-1,0\}\\
$H_{d_i}$ & $(1,2,\frac{1}{2}) $ & \{0,1,-1\} & \{1,-1,0\}\\
$\eta_{i}$ & $(1,1,0) $ & \{0,1,-1\} & \{1,-1,0\}\\
\hline 
$T_{L}, T_{R} $ & $(3,1,\frac{2}{3}) $ & 0 & 0 \\
$B_{L}, B_{R} $ & $(3,1,-\frac{1}{3}) $ & 0 & 0 \\
$E_{L}, E_{R} $ & $(1,1,-1) $ & 0 & 0 \\
\hline
\hline
\end{tabular}
\end{center}
\caption{The SM and $G_F$ charges of various fermions and scalars of the model. Here, $i=1,2,3$ denote three generations and their respective charges under new $U(1)$ are represented as $\{q_1,q_2,q_3\}$.}
\label{tab:fields}
\end{table}


\subsection{Charged fermion masses}
The most general renormalizable couplings between the fermions and scalars invariant under the SM gauge symmetry and $G_F$ can be written as
\beqa \label{LY}
-{\cal L}_Y &=& {y_u}_i\,\overline{Q_{L}}_i\, {H_u}_i\, T_R\, + {y_u^{\prime}}_i\,\overline{T_L}\, \eta^*_i\, u_{R i}\,+\,{y_d}_i\,\overline{Q_{L}}_i\, {H_d}_i\, B_R\, + {y_d^{\prime}}_i\,\overline{B_L}\, \eta^*_i\, d_{R i}\, \nonumber \\
& + & {y_e}_i\,\overline{L_L}_i\, {H_d}_i\, E_R\, + {y_e^{\prime}}_i\,\overline{E_L}\, \eta^*_i\, e_{R i}\, + \, {\rm h.c.}\,.\eeqa
The direct couplings between two SM fermions and Higgs are forbidden by $G_F$. The masses of vectorlike fermions   are parametrized as 
\be \label{LM}
-{\cal L}_m = m_T\, \overline{T_L}\,T_R\, + \,m_B\, \overline{B_L}\,B_R\, + \,m_E\, \overline{E_L}\,E_R\, + \, {\rm h.c.}\,.\ee

The spontaneous breaking of $G_F$ through the Vacuum Expectation Values (VEVs) of $\eta_i$ and ${H_{ui,di}}$ gives rise to $4 \times 4$ mass matrices for the charged fermions identical to the one given in Eq. (\ref{M}). Explicitly,
\be \label{M_expl}
{\cal M}_{u,d,e} = \left( \ba{cc} 0 & \left(\mu_{u,d,e}\right)_{3 \times 1} \\ \left(\mu^\prime_{u,d,e}\right)_{1 \times 3} & m_{T,B,E} \ea \right)\,,\ee
where 
\be \label{mu_expl}
{\mu_u}_i = {y_u}_i {v_u}_i\,,~~{\mu_d}_i = {y_d}_i {v_d}_i\,,~~{\mu_e}_i = {y_e}_i {v_d}_i\,,\ee
\be \label{mup_expl}
{\mu^\prime_u}_i = {y^\prime_u}_i \langle \eta_i \rangle\,,~~{\mu^\prime_d}_i = {y^\prime_d}_i \langle \eta_i \rangle\,,~~{\mu^\prime_e}_i = {y^\prime_e}_i \langle \eta_i \rangle\,,\ee
and ${v_u}_i \equiv \langle {H_u}_i\rangle$, ${v_d}_i \equiv \langle {H_d}_i\rangle$. The repeated indices do not imply summation in the above. The above VEVs are chosen such that they break $G_F$ entirely and $SU(2)_L \times U(1)_Y$ into $U(1)_{\rm em}$. The most general potential of the model is given in Appendix \ref{app:scalar}. It is shown that there exist a large number of parameters which may allow the desired solution for the VEVs although we do not study the potential minimization in detail. The effective $3 \times 3$ mass matrix in each sector, analogous to Eq. (\ref{M_eff}), can be written as
\be \label{Meff_0}
M_{u,d,e}^{(0)} \equiv - \frac{1}{m_{T,B,E}}\, \mu_{u,d,e}\, \mu^\prime_{u,d,e}\,.\ee
The above matrices are of rank one and give masses to the third generation charged fermions.

Following the procedure outlined in the previous section and Eq. (\ref{M_eff_corr}), the 1-loop corrected effective $3 \times 3$ mass matrices for the charged fermions can be obtained as
\be \label{Mf}
M_f =  \delta M_f + M_f^{(0)}\,,\ee 
where $f=u,d,e$ and the second term is written using Eq. (\ref{Meff_0}). In comparison to general result given in Eq. (\ref{M_eff_corr}), note that the second term is same as the tree level effective mass matrix. The loop corrections do not modify the parameters $\mu_i$, $\mu^\prime_i$ and $m_F$ as the vectorlike fermions are neutral under the $G_F$. $\delta M_f$ contains 1-loop correction induced by the gauge bosons of both the $U(1)$s. Using the given $U(1)_{1,2}$ charges and Eq. (\ref{del_calM_33}), we find
\beqa \label{delMf}
\delta M_f &=& \frac{N_f g_1^2}{4 \pi^2} \left(b_0[M_{Z_1}^2,m_{f3}^2] - b_0[M_{Z_1}^2,m_F^2]\right)\,\left(\ba{ccc} 0 & 0 & 0\\0 & \left(M_f^{(0)}\right)_{22} &  -\left(M_f^{(0)}\right)_{23} \\
0 & -\left(M_f^{(0)}\right)_{32} &  \left(M_f^{(0)}\right)_{33} \ea \right) \nonumber \\
& + & \frac{N_f g_2^2}{4 \pi^2} \left(b_0[M_{Z_2}^2,m_{f3}^2] - b_0[M_{Z_2}^2,m_F^2]\right)\,\left(\ba{ccc}  \left(M_f^{(0)}\right)_{11} &  -\left(M_f^{(0)}\right)_{12} & 0 \\
-\left(M_f^{(0)}\right)_{21} &  \left(M_f^{(0)}\right)_{22} & 0 \\ 0 & 0 & 0 \ea \right)\,. \eeqa
Here, $g_{i}$ is the gauge coupling and $M_{Z_i}$ is the mass of the gauge boson of $U(1)_i$. $N_f = 3$ ($N_f=1$) for $f=u,d$ ($f=e$) is the factor due to color degrees of freedom. $m_{f3}$ and $m_F$ are masses of the third generation fermion and vectorlike fermion, respectively, in each sector.

The charged fermion mass matrix $M_f$ defined in Eq. (\ref{Mf}) along with expressions (\ref{Meff_0}) and (\ref{delMf}) gives the 1-loop  corrected mass matrix in the model. Each of the 1-loop generated contributions in $\delta M_f$ are of rank one and induce masses for the first and second generation fermions individually. The hierarchy among the masses of the lighter generations can be arranged by choosing $M_{Z_1} \ll M_{Z_2}$. This suppression is common for all the charged fermions.

Note that in determining Eq. (\ref{delMf}), we have considered 1-loop corrections only from the gauge boson loops. In the present model, radiative corrections also arise from the emission and absorption of scalar bosons in the loop. Such a contribution requires mixing between the scalar fields $\eta_i$ and $H_{ui,di}$ as they exclusively couple to the right and left chiral fermions, respectively. These corrections can be made suppressed by choosing the scalar masses much greater than the $M_{Z_1}$ \cite{Weinberg:2020zba} and/or by assuming small mixing between $\eta_i$ and $H_{ui,di}$ fields \cite{Jana:2021tlx}. We assume that such suppression is arranged and do not consider the radiative corrections from the scalar bosons. 

We have also assumed that there are no kinetic mixings between different $U(1)$. Presence of such a mixing, for example $\epsilon {F_1}_{\mu \nu} {F_2}^{\mu \nu}$ where $F_{1,2}$ are the field strengths of $Z_{1,2}$ bosons, can allow the first-generation fermions to receive masses from the 1-loop diagrams involving $Z_1$ boson. This can be understood from the fact that an effective coupling of order $\epsilon g_1$ gets induced between the first-generation fermions and $Z_1$ when the kinetic mixing term is rotated away to obtain the physical gauge bosons (see for example \cite{Babu:1996vt}). Both the first and second generations receive mass from $Z_1$ loop in this case, however, the former is suppressed by an additional factor of $\epsilon^2$. Therefore, the presence of kinetic mixing is not expected to spoil the hierarchies between the first two generations in this model if $\epsilon \ll M_{Z_1}/M_{Z_2}$.

\subsection{Neutrino masses}
Although our main aim is to explain charged fermion mass hierarchies, we also comment on the possibility of neutrino mass generation within this model. Most simply, the naturally small neutrino masses can be accommodated by introducing three RH neutrinos, singlet under both the SM gauge symmetry and $G_F$, and allowing Majorana masses for them. The gauge invariant renormalizable interactions can be parametrized as
\be \label{L_nu}
-{\cal L}_\nu = {y_D}_{ij}\,\overline{L_L}_i\,{H_u}_i\,{\nu_R}_j + \frac{1}{2} {M_R}_{ij}\,\nu_{R i}^T C^{-1} \nu_{R j} + {\rm h.c.}\,. \ee 
Electroweak symmetry breaking gives rise to the Dirac neutrino mass matrix, ${M_D}_{ij} ={ y_D}_{ij} {v_u}_i$. For $M_R \gg M_D$, the usual type I seesaw mechanism \cite{Minkowski:1977sc,Yanagida:1979as,Mohapatra:1979ia} can be realized and the light neutrino mass matrix is given by
\be \label{M_nu}
M_{\nu} = - M_D\,M_R^{-1}\, M_D^T\,. \ee
Unlike the charged fermions, all the three neutrino masses can arise at the tree level in general. This is a welcome feature as the inter-generation hierarchies in the neutrino masses are not as strong as that in the charged fermion masses. All the elements of the Dirac neutrino Yukawa coupling matrix $y_D$ can be of ${\cal O}(1)$ leading to an anarchic structure for $M_\nu$. This, in turn, can explain the relatively feeble hierarchy in the neutrino masses and the large mixing in the lepton sector \cite{Hall:1999sn,deGouvea:2012ac}.

\section{Example solutions}
\label{sec:solutions}
We investigate the viability of the proposed framework in reproducing charged fermion masses and quark mixing by finding numerical values for the parameters $\mu_{f i}$, $\mu^\prime_{f i}$, $m_T$, $m_B$, $m_E$ and $M_{Z_{1,2}}$. It can be noted from Eq. (\ref{LY}) that $y_{u i}$, $y^\prime_{u i}$, $y^\prime_{d i}$, $y_{ei}$ and $y^\prime_{ei}$ can be chosen real by rotating away their phases through redefinitions of the various quarks and lepton fields. Similarly, one of the $y_{di}$ can also be made real. An analogous treatment in Eq. (\ref{LM}) leads to real $m_T$, $m_B$ and $m_E$. Moreover, we assume that all the VEVs are real. Altogether, this implies 25 real parameters (real $\mu_{u i}$, $\mu^\prime_{ui}$, $\mu_{d3}$, $\mu^\prime_{di}$, $\mu_{ei}$, $\mu^\prime_{ei}$, $m_T$, $m_B$, $m_E$, $M_{Z_1}$, $M_{Z_2}$ and complex $\mu_{d1}$, $\mu_{d2}$) which can be used to determine 13 observables (9 charged fermion masses, 3 angles and a Dirac CP phase of the quark mixing matrix). The number of parameters is greater than the number of observables and hence one expects to get viable solutions. Nevertheless, considering that masses and mixing observables are complex non-linear functions of the input parameters and the latter are expected to take not-so-hierarchical values in the present model, it is not obvious that viable solutions would exist.

We determine the 25 real parameters of the underlying framework through the usual $\chi^2$ function minimization technique. The $\chi^2$ function (see for example \cite{Mummidi:2021anm} for the definition and details) consists of 13 observables, the mean values and standard deviations of which are listed in Table \ref{tab:input}. 
\begin{table}[t]
\begin{center}
\begin{tabular}{cccccc} 
\hline
\hline
~~Observable~~&~~Value~~&~~Observable~~&~~Value~~\\
 \hline
$m_u$ & $1.27\pm 0.50$ MeV & $m_e$ & $0.487 \pm 0.049$ MeV \\
$m_c$ & $0.619 \pm 0.084$ GeV & $m_\mu$ &  $1.027\pm 0.103$ MeV \\
$m_t$ & $171.7\pm 3.0$ GeV & $m_\tau$ & $1.746 \pm 0.174$ GeV\\
$m_d$ & $2.90 \pm 1.24$ MeV & $|V_{us}|$ & $0.22500 \pm 0.00067$ \\
$m_s$ & $0.055 \pm 0.016$ GeV & $|V_{cb}|$ & $0.04182 \pm 0.00085$  \\
$m_b$ & $2.89 \pm 0.09$ GeV & $|V_{ub}|$ & $0.00369 \pm 0.00011$ \\
 &  & $J_{\rm CP}$ & $(3.08 \pm 0.15)\times 10^{-5}$ \\
\hline
\hline
\end{tabular}
\end{center}
\caption{Values of charged fermion masses and CKM parameters extrapolated at $M_Z$ used in the fits to obtain the example solutions. The values of the charged fermion masses and quark mixing parameters are taken from \cite{Xing:2007fb} and \cite{ParticleDataGroup:2020ssz}, respectively.}
\label{tab:input}
\end{table}
For definiteness on the mass scale of new physics, we take three different values for $M_{Z_1}$ and obtain a benchmark solution for each. The optimized values of the remaining parameters are listed in Table \ref{tab:sol} for each solution.
\begin{table}[t]
\begin{center}
\begin{tabular}{cccc} 
\hline
\hline
~~Parameters~~&~~Solution 1 (S1)~~&~~Solution 2 (S2)~~&~~Solution 3 (S3)~~\\
\hline
$M_{Z_1}$ & $10^4$ & $10^6$ & $10^8$ \\
$M_{Z_2}$ & $2.8708 \times 10^{5}$  & $ 1.4470\times 10^7$  & $1.9110\times 10^9 $\\
$m_T$ & $1.1000 \times 10^{4}$  & $1.1000\times 10^6 $  & $1.1003 \times 10^8$\\
$m_B$ & $3.0754\times 10^{5}$  & $2.4839\times 10^7 $  & $1.4015\times 10^9 $\\
$m_E$ & $2.8128\times 10^{5} $ & $4.9462\times 10^7$ & $6.7820\times 10^8 $ \\
\hline
$\mu_{u1}$ & $1.8023\times 10^1 $ & $-1.6702\times 10^1$ & $1.5410 \times 10^1$\\
$\mu_{u2}$ & $3.0901 \times 10^2$ & $3.0969\times 10^2 $ & $-0.6860 $ \\
$\mu_{u3}$ & $1.4763$ & $-0.5133$ & $2.8599 \times 10^2$\\
\hline
$\mu^\prime_{u1}$ & $-3.9573\times 10^{3}$ & $3.3923\times 10^5 $ & $-3.8518\times 10^7 $\\
$\mu^\prime_{u2}$ & $3.5446 \times 10^{3}$ & $-3.9343 \times 10^5 $ & $-3.7762\times 10^7$ \\
$\mu^\prime_{u3}$ & $-2.8507\times 10^{3}$ & $29653\times 10^5 $ & $-3.7718\times 10^7$\\
\hline
$\mu_{d1}$ & $2.3809\times 10^{1} + i\ 6.7460 $ & $-1.0402 \times 10^{1} + i\ 6.7204  $ & $4.6117 + i\ 4.3534  $ \\
$\mu_{d2}$ & $1.4422 \times 10^2+ i\ 2.6938 \times 10^1$ & $2.7279  \times 10^{1} - i\ 1.2285 \times 10^{2}$ & $-0.1868- i\ 0.8509 $ \\
$\mu_{d3}$ & $5.6345$ & $-2.1306$ & $1.0439 \times 10^2$ \\
\hline
$\mu^\prime_{d1}$ & $-6.1152\times 10^2 $ & $8.8600\times 10^4 $ & $-2.2724\times 10^7$\\
$\mu^\prime_{d2}$ & $ 3.7385\times 10^3 $ & $3.4743\times 10^5$ & $2.4618\times 10^7$ \\
$\mu^\prime_{d3}$ & $-7.5116 \times 10^2 $ & $1.2636 \times 10^5$ & $-1.9748\times 10^7 $ \\
\hline
$\mu_{e1}$ & $ 8.1306 \times 10^1 $ & $-7.6362 \times 10^1 $ & $-0.9180$\\
$\mu_{e2}$ & $2.7874 \times 10^1 $ & $-1.5295\times 10^{2} $ & $-9.6383$ \\
$\mu_{e3}$ & $-1.1628$      & $3.2793$    & $-2.2318\times 10^1 $ \\
\hline
$\mu^\prime_{e1}$ & $-1.2295 \times 10^3 $ & $-1.9688\times 10^5$    & $2.7449\times 10^7 $\\
$\mu^\prime_{e2}$ & $-3.9625 \times 10^3 $ & $-3.8735 \times 10^5 $ & $-3.5794\times 10^7 $ \\
$\mu^\prime_{e3}$ & $3.9792 \times 10^3 $  & $2.9732 \times 10^4 $ & $2.0714\times 10^7 $\\
\hline
\hline
\end{tabular}
\end{center}
\caption{Optimized values of various input parameters obtained for three example solutions. All the values are in GeV.}
\label{tab:sol}
\end{table}
All the solutions provide an excellent fit to the observables and reproduce their central values with a total $\chi^2 \ll 1$ in all three cases. We do not include neutrino masses and lepton mixing in the fit as they are given by an entirely different set of parameters, see Eq. (\ref{M_nu}), which are arbitrary. The latter can be chosen to reproduce viable neutrino masses and mixing parameters.

The example solutions given in Table \ref{tab:sol} indicate that the model can reproduce a realistic charged fermion spectrum irrespective of the scales of $U(1)_{1,2}$ breaking. As discussed earlier, the mass gap between $Z_1$ and $Z_2$ is determined by the mass hierarchies between the first and second generation fermions and one finds $M_{Z_2}^2/M_{Z_1}^2 \simeq {\cal O}(10^2)$, however, the absolute scale of $M_{Z_1}$ is not fixed. We impose $M_{Z_1} \le m_T, m_B, m_E$ while performing the numerical fits. It is found that $m_T \ll m_B, m_E$ is required in order to generate $m_t \gg m_b, m_\tau$. The parameters $\mu_{fi}$, $f=u,d,e$, are generated through electroweak symmetry breaking, see Eq. (\ref{mu_expl}), for which we impose $|y_{fi}|<\sqrt{4 \pi}$ and $v_{u i},v_{d i} < 174$ GeV. As a result, all the $\mu_{fi}$ are found to be of ${\cal O}(100)$ GeV or less. On the other hand, $\mu^\prime_{fi}$ are induced by the VEVs of SM singlet but $U(1)_{1,2}$ charged fields and their determined values are close to the breaking scale of $U(1)_{1,2}$. The most noteworthy feature of each of the solutions obtained in Table \ref{tab:sol} is that there are no large hierarchies within the values of various $\mu_{fi}$ or $\mu^\prime_{fi}$. This implies that the dimensionless parameters of the underlying framework can be of the same magnitude. Despite this, the observed difference of five orders of magnitude between the masses of the first and third generation fermions is achieved through the radiative mass generation mechanism.

\section{Phenomenological aspects}
\label{sec:pheno}
The model has rich phenomenological implications due to the presence of two flavourful $U(1)$ gauge symmetries and additional vectorlike pair of quarks and leptons. For all the solutions listed in the previous section, one finds $Z_1$  as the lightest among all the new particles. We, therefore, discuss various constraints on $Z_1$ boson arising from the processes involving quark and lepton Flavour Changing Neutral Currents (FCNCs). 

In the physical basis of quarks and leptons, the couplings of $Z_{1,2}$ can be determined from Eqs. (\ref{L_gauge_2},\ref{U_block}) as:
\be \label{Z_couplings}
-{\cal L}_{Z_{1,2}} =  \sum_{k=1,2}\,g_{k}\, \left(\left(X^{(k)}_{f_L}\right)_{ij}\, \overline{f_{L i}}\,\gamma^\mu f_{L j} + \left(X^{(k)}_{f_R}\right)_{ij}\, \overline{f_{R i}}\,\gamma^\mu f_{R j}\right) Z_{k \mu}\,, \ee
where $f=u,d,e$. The $3 \times 3$ coupling matrices are given by
\be \label{X}
X^{(k)}_{f_L} = U_{f_L}^\dagger\, q^{(k)}_{f L}\, U_{f_L}\,,\ee
and similar expression for $X^{(k)}_{f_R}$ can be obtained by replacing $L \to R$. In the present framework, we have $q^{(1)}_{f L} = q^{(1)}_{f_R} = {\rm Diag.}(0,1,-1)$  and $q^{(2)}_{f L} = q^{(2)}_{f_R} = {\rm Diag.}(1,-1,0)$ for all $f$ as specified earlier. The unitary matrices $U_{f_L}$ and $U_{f_R}$ are obtained from diagonalizing the 1-loop corrected mass matrices $M_f$ such that $U_{f_L}^\dagger M_f U_{f_R} = {\rm Diag.}(m_{f_1},m_{f_2},m_{f_3})$.  Since the underlying $U(1)$ symmetries are non-universal, $X^{(k)}_{f_L}$ and $X^{(k)}_{f_R}$ are non-diagonal in general. This can lead to large flavour violations both in the quark and lepton sectors.  The numerical values of various $X^{(k)}_{f_L}$ and $X^{(k)}_{f_R}$ for one of the benchmark solutions are given in Appendix \ref{app:Xvalues} for reference.

\subsection{Quark flavour violation}
Due to its non-vanishing off-diagonal couplings with the quarks, the $Z_1$ boson contributes to the meson-antimeson mixing at the tree level itself. To estimate these contributions for $K^0-\overline{K}^0$, $B_d^0-\overline{B}_d^0$, $B_s^0-\overline{B}_s^0$ and $D^0-\overline{D}^0$, we follow the effective operator based analysis (see for example \cite{UTfit:2007eik}) and parametrize the new contributions in terms of the well-known Wilson coefficients (WCs). Subsequently, we use the limits on these coefficients obtained from a fit to experimental data by UTFit collaboration \cite{UTfit:2007eik} to derive constraints on the mass scale of $Z_1$ boson.

For $K^0-\overline{K}^0$ mixing, the effective Hamiltonian for $\Delta S=2$ is written as $H_{\rm eff} = \sum_{i=1}^5 C_K^i\,Q_i + \sum_{i=1}^3 \tilde{C}_K^i \tilde{Q}_i$ where explicit form of operators are listed in \cite{UTfit:2007eik,Ciuchini:1998ix}. When $Z_1$ is integrated out, its contribution to the Wilson coefficients $C_K^i$ and $\tilde{C}_K^i$ is obtained, at the scale $\mu=M_{Z_1}$,  as \cite{Smolkovic:2019jow}
\be \label{C_K}
C_K^1=\frac{g_1^2}{M_{Z_1}^2}\left[\left(X^{(1)}_{d_L}\right)_{12}\right]^2\,,~~\tilde{C}_K^1=\frac{g_1^2}{M_{Z_1}^2}\left[\left(X^{(1)}_{d_R}\right)_{12}\right]^2\,,~~C_K^5=-4\frac{g_1^2}{M_{Z_1}^2}\left(X^{(1)}_{d_L}\right)_{12} \left(X^{(1)}_{d_R}\right)_{12}\,. \ee 
The remaining $C_K^i$ and $\tilde{C}_K^i$ have vanishing values at $\mu=M_{Z_1}$. For a precise comparison with the experimental values, these coefficients need to be evolved from $\mu=M_{Z_1}$ to $\mu=2$ GeV. We perform such running using the Renormalization Group Equations (RGE) given in \cite{Ciuchini:1998ix}. It is found that RGE effects induce non-zero $C_K^4$ while the $C^{2,3}_K$ and $\tilde{C}^{2,3}_K$ have vanishing values at the low scale. The RGE evolved values are listed in Table \ref{tab:meson_WC} and compared with the corresponding experimentally allowed ranges extracted from a fit performed by the UTFit collaboration \cite{UTfit:2007eik}.
\begin{table}[t]
\begin{center}
\begin{tabular}{ccccc} 
\hline
\hline
~~Wilson coefficient ~~&~~Allowed range~~&~~S1~~&~~S2~~&~~S3~~\\
 \hline
Re$C_K^1$ & $[-9.6,9.6]\times 10^{-13}$ & \rb{$-9.5\times 10^{-10}$} & \gb{$-5.4 \times 10^{-14} $} & \gb{$6.2\times 10^{-18}$}\\
Re$\tilde{C}_K^1$ & $[-9.6,9.6]\times 10^{-13}$ & \rb{$-1.6\times 10^{-9}$}& \gb{$-1.6\times 10^{-13}$} & \gb{$2.8\times 10^{-17}$}\\
Re$C_K^4$ & $[-3.6,3.6]\times 10^{-15}$ & \rb{$6.2\times 10^{-9}$} & \rb{$5.0\times 10^{-13}$} & \gb{$-7.5 \times 10^{-17}$} \\
Re$C_K^5$ & $[-1.0,1.0]\times 10^{-14}$ & \rb{$5.4\times 10^{-9}$} & \rb{$4.2\times 10^{-13}$} & \gb{$-5.9\times 10^{-17}$}\\
Im$C_K^1$ & $[-9.6,9.6]\times 10^{-13}$ & \gb{$5.9\times 10^{-25}$}& \gb{$9.5\times 10^{-30}$} & \gb{$1.7\times 10^{-33}$}\\
Im$\tilde{C}_K^1$ & $[-9.6,9.6]\times 10^{-13}$ &\gb{$-1.0 \times 10^{-24}$} & \gb{$-3.8\times 10^{-29}$} & \gb{$3.9\times 10^{-31}$}\\
Im$C_K^4$ &  $[-1.8,0.9]\times 10^{-17}$  & \gb{$9.5\times 10^{-26}$}  & \gb{$1.5\times 10^{-29}$} &\gb{$-5.3\times 10^{-31}$} \\
Im$C_K^5$ & $[-1.0,1.0]\times 10^{-14}$ &\gb{$8.3\times 10^{-26}$} & \gb{$1.3\times 10^{-29}$} & \gb{$-4.2\times 10^{-31}$}\\
\hline
$|C_{B_d}^1|$ & $<2.3\times 10^{-11}$ & \gb{$1.6\times 10^{-12}$} & \gb{$9.9\times 10^{-18}$} & \gb{$5.8 \times 10^{-22}$} \\
$|\tilde{C}_{B_d}^1|$ & $<2.3\times 10^{-11}$  & \gb{$2.9\times 10^{-12}$} & \gb{$3.8\times 10^{-18}$} & \gb{$1.0\times 10^{-18}$}\\
$|C_{B_d}^4|$ &  $<2.1\times 10^{-13}$ &  \rb{$5.1 \times 10^{-12}$}   & \gb{$1.6\times 10^{-17}$} & \gb{$6.7\times 10^{-20}$}\\
$|C_{B_d}^5|$ & $<6.0\times 10^{-13}$  & \rb{$9.1 \times 10^{-12}$} & \gb{$2.6\times 10^{-17}$} & \gb{$1.0\times 10^{-19}$}\\
\hline
$|C_{B_s}^1|$ & $< 1.1 \times 10^{-9}$ & \gb{$8.3 \times 10^{-11}$} & \gb{$2.8\times 10^{-15}$} & \gb{$3.0\times 10^{-19}$}\\
$|\tilde{C}_{B_s}^1|$ & $< 1.1 \times 10^{-9}$ & \gb{$2.0\times 10^{-10}$} & \gb{$5.8\times 10^{-14}$} & \gb{$4.2\times 10^{-17}$}\\
$|C_{B_s}^4|$ & $< 1.6 \times 10^{-11}$ & \rb{$3.1 \times 10^{-10}$}  & \gb{$3.3\times 10^{-14}$} & \gb{$9.8\times 10^{-18}$} \\
$|C_{B_s}^5|$ & $< 4.5 \times 10^{-11}$ & \rb{$5.5 \times 10^{-10}$} & \gb{$5.4\times 10^{-14}$} & \gb{$1.5 \times 10^{-17}$}\\
\hline
$|C_D^1|$ & $<7.2 \times 10^{-13}$ & \rb{$2.0\times 10^{-10}$} & \gb{$2.9\times 10^{-15}$} & \gb{$6.5\times 10^{-19}$}\\
$|\tilde{C}_D^1|$ & $<7.2 \times 10^{-13}$ & \rb{$3.5 \times 10^{-9}$} & \gb{$2.7\times 10^{-13}$} & \gb{$2.8\times 10^{-17}$}\\
$|C_D^4|$ & $<4.8\times 10^{-14}$ & \rb{$3.2\times 10^{-9}$} & \rb{$1.1\times 10^{-13}$} & \gb{$1.7\times 10^{-17}$} \\
$|C_D^5|$ & $<4.8 \times 10^{-13}$ & \rb{$3.7\times 10^{-9}$} & \gb{$1.2\times 10^{-13}$} & \gb{$1.9\times 10^{-17}$}\\
\hline
\hline
\end{tabular}
\end{center}
\caption{Strength of various Wilson coefficients relevant for meson-antimeson mixing estimated for three example solutions and corresponding experimentally allowed range at $95\%$ confidence level. For the later we use the results from \cite{UTfit:2007eik}. All the values are in ${\rm GeV}^{-2}$. The values highlighted in green (red) are allowed (excluded) by experimental limits.}
\label{tab:meson_WC}
\end{table}

In case of  $B_q-\overline{B}_q^0$ $(q=d,s)$ mixing, the relevant WCs at $\mu=M_{Z_1}$ are \cite{Smolkovic:2019jow}
\be \label{C_Bd}
C_{B_d}^1=\frac{g_1^2}{M_{Z_1}^2}\left[\left(X^{(1)}_{d_L}\right)_{13}\right]^2\,,~~\tilde{C}_{B_d}^1=\frac{g_1^2}{M_{Z_1}^2}\left[\left(X^{(1)}_{d_R}\right)_{13}\right]^2\,,~~C_{B_d}^5=- 4\frac{g_1^2}{M_{Z_1}^2}\left(X^{(1)}_{d_L}\right)_{13} \left(X^{(1)}_{d_R}\right)_{13}\,,\ee
and 
\be \label{C_Bs}
C_{B_s}^1=\frac{g_1^2}{M_{Z_1}^2}\left[\left(X^{(1)}_{d_L}\right)_{23}\right]^2\,,~~\tilde{C}_{B_s}^1=\frac{g_1^2}{M_{Z_1}^2}\left[\left(X^{(1)}_{d_R}\right)_{23}\right]^2\,,~~C_{B_s}^5=- 4\frac{g_1^2}{M_{Z_1}^2}\left(X^{(1)}_{d_L}\right)_{23} \left(X^{(1)}_{d_R}\right)_{23}\,.\ee
These coefficients are run down to $\mu = M_b = 4.6$ GeV following \cite{Becirevic:2001jj}. Similarly, for the charm mixing governing the $D^0-\overline{D}^0$ oscillations, the WCs at the scale $M_{Z_1}$ are given by 
\be \label{C_D}
C_D^1=\frac{g_1^2}{M_{Z_1}^2}\left[\left(X^{(1)}_{u_L}\right)_{12}\right]^2\,,~~\tilde{C}_D^1=\frac{g_1^2}{M_{Z_1}^2}\left[\left(X^{(1)}_{u_R}\right)_{12}\right]^2\,,~~C_D^5=- 4\frac{g_1^2}{M_{Z_1}^2}\left(X^{(1)}_{u_L}\right)_{12} \left(X^{(1)}_{u_R}\right)_{12}\,.\ee
They are also evolved to the relevant low scale $\mu = 2.8$ GeV using the RGE equations given in \cite{UTfit:2007eik}.

We list the values of all the WCs at their relevant hadronic scales for the three benchmark solutions and compare them with the corresponding experimental limits in Table \ref{tab:meson_WC}. It is noticed that constraints from meson-antimeson mixings put a strong limit on the mass of $Z_1$ boson since the latter typically has ${\cal O}(1)$ off-diagonal couplings with the quark flavours, see Appendix \ref{app:Xvalues} for example. It can be seen that both the solutions S1 and S2 are disfavoured implying $M_{Z_1} > 10^3$ TeV for phenomenologically consistent solutions. This also implies that the model cannot account for the neutral current $B$-anomalies which generically require $M_{Z_1} \le 2$ TeV \cite{DiLuzio:2017fdq,Allanach:2019mfl}.

\subsection{Lepton flavour violation}
The flavourful $Z_1$ mediate charged lepton flavour violating processes like $\mu$ to $e$ conversion in nuclei and $l_i \to 3 l_j$ at tree level. The processes like $l_i \to l_j \gamma$ arise at 1-loop through $Z_1$ and the charged leptons in the loop. In this subsection, we estimate the constraints on $Z_1$ from all these processes. 

In the field of nucleus, the muons can undergo transition to electrons through flavour violating coupling of $\mu$ and $e$ with $Z_1$ boson. The strongest limit on such a process has been obtained by SINDRUM II experiment which uses $^{197}$Au nucleus \cite{SINDRUMII:2006dvw}. The branching ratio for this process computed in \cite{Kitano:2002mt} is given by
\be \label{mu2e}
{\rm BR}[\mu \to e] = \frac{2 G_F^2}{\omega_{\rm capt}}\,(V^{(p)})^2\, \left(|g^{(p)}_{LV}|^2 + |g^{(p)}_{RV}|^2\right)\,, \ee
where $V^{(p)}$ is an integral involving proton distribution in a given nucleus, $\omega_{\rm capt}$ is muon capture rate by the nucleus and 
\be \label{gLV}
g^{(p)}_{LV,RV} = 2 g^{(u)}_{LV,RV} + g^{(d)}_{LV,RV}\,.\ee
For $Z_1$ mediated contributions and $M_{Z_1} \gg m_\mu$, the above couplings are given by \cite{Smolkovic:2019jow}
\be \label{gLV_Z1}
g^{(f)}_{LV} \sim \frac{\sqrt{2}}{G_F} \frac{g_1^2}{M_{Z_1}^2}\,\left(X^{(1)}_{e_L}\right)_{12} \frac{1}{2}\left[\left(X^{(1)}_{f_L}\right)_{11} + \left(X^{(1)}_{f_R}\right)_{11} \right]\,\ee
with $f=u,d$. Similarly, $g^{(f)}_{RV} $ is given by replacement $L \leftrightarrow R$ in the above expression. Substituting Eqs. (\ref{gLV_Z1},\ref{gLV}) in (\ref{mu2e}) and using $V^{(p)}= 0.0974\, m_\mu^{5/2}$, $\omega_{\rm capt} = 13.07 \times 10^{6}\,{\rm s}^{-1}$ for $^{197}$Au from \cite{Kitano:2002mt}, we estimate ${\rm BR}[\mu \to e]$ for the obtained solutions and list them in Table \ref{tab:LFV}. We also give the latest experimental limit on ${\rm BR}[\mu \to e]$  in the same table for comparison.
\begin{table}[t]
\begin{center}
\begin{tabular}{ccccc} 
\hline
\hline
~~LFV observable~~&~~Limit~~&~~S1~~&~~S2~~&~~S3~~\\
 \hline
${\rm BR}[\mu \to e]$ & $< 7.0 \times 10^{-13}$ & \rb{$7.2 \times 10^{-7}$} & \gb{$4.0 \times 10^{-15}$} & \gb{$3.7 \times 10^{-22}$}\\
\hline
${\rm BR}[\mu \to 3e]$ & $< 1.0 \times 10^{-12}$  & \rb{$7.9 \times 10^{-9}$} & \gb{$6.0 \times 10^{-17}$} & \gb{$2.5 \times 10^{-25}$}\\
${\rm BR}[\tau \to 3\mu]$ & $< 2.1 \times 10^{-8}$ & \rb{$2.3\times 10^{-8}$} & \gb{$1.7\times 10^{-18}$}& \gb{$1.1 \times 10^{-24}$}\\
${\rm BR}[\tau \to 3 e]$ & $< 2.7 \times 10^{-8}$ & \gb{$9.2\times 10^{-11}$} & \gb{$6.5\times 10^{-19}$}& \gb{$4.2 \times 10^{-28}$}\\
\hline
${\rm BR}[\mu \to e \gamma]$ & $< 4.2 \times 10^{-13}$ & \rb{$2.0\times 10^{-11}$} & \gb{$1.3\times 10^{-19}$}& \gb{$3.5 \times 10^{-27}$}\\
${\rm BR}[\tau \to \mu \gamma]$ &  $< 4.4 \times 10^{-8}$ & \gb{$9.3\times 10^{-13}$}& \gb{$7.1\times 10^{-20}$}& \gb{$3.1 \times 10^{-27}$}\\
${\rm BR}[\tau \to e \gamma]$ &  $< 3.3 \times 10^{-8}$ & \gb{$3.9\times 10^{-13}$} & \gb{$2.1\times 10^{-21}$}& \gb{$4.0 \times 10^{-29}$}\\
\hline
\hline
\end{tabular}
\end{center}
\caption{Estimated values of various charged lepton flavour violating observables for the three benchmark solutions and the present experimental limits at $90 \%$ confidence level. The latter are taken from \cite{Calibbi:2017uvl}. The values highlighted in green (red) are allowed (excluded) by experimental limits.}
\label{tab:LFV}
\end{table}

Next, we estimate the branching ratios of $\mu \to 3 e$, $\tau \to 3 \mu$ and $\tau \to 3 e$ following \cite{Heeck:2016xkh,Smolkovic:2019jow}. The relevant decay width, estimated neglecting sub-leading terms proportional to $m_{l_j}$, is given by
\beqa \label{lto3l}
\Gamma[l_i \to 3 l_j] &\simeq& \frac{g_1^4 m_{l_i}^5}{768 \pi^3 M_{Z_1}^4}\,\left[ 4 {\rm Re}\left( \left(X_{eV}\right)_{ji} \left(X_{eA}\right)_{ji} \left(X_{eV}\right)^*_{jj} \left(X_{eA}\right)^*_{jj} \right) \right. \nonumber \\
&+& \left. 3 \left( \left| \left(X_{eV}\right)_{ji}\right|^2 + \left| \left(X_{eA}\right)_{ji}\right|^2 \right) \left( \left| \left(X_{eV}\right)_{jj}\right|^2 + \left| \left(X_{eA}\right)_{jj}\right|^2 \right) \right]\,,
\eeqa
where
\be \label{X_VA}
X_{eV,eA} = \frac{1}{2} \left(X^{(1)}_{e_L}  \pm X^{(1)}_{e_R}\right)\,, \ee
are couplings for vector and axial-vector currents, respectively.  Using the above expression, the evaluated numbers for ${\rm BR}[l_i \to 3 l_j]$ are given in Table \ref{tab:LFV} along with the latest limits from experiments.

Unlike the previous decays, the decays like $l_i \to l_j \gamma$ arise at 1-loop level. Nevertheless, we estimate these decays considering relatively strong limits on ${\rm BR}[\mu \to e \gamma]$. The corresponding decay width is given by \cite{Lavoura:2003xp}
\be \label{muegamma}
\Gamma[l_i \to l_j \gamma] = \frac{\alpha g_1^4}{4 \pi}\,\left(1-\frac{m_{l_j}^2}{m_{l_i}^2}\right)^3\,\frac{m_{l_i}^4}{M_{Z_1}^4}\,m_{l_i}\,\left( |c^\gamma_L|^2+|c_R^\gamma|^2\right)\,.\ee
 Here, 
 \beqa \label{c_gamma}
 c_L^\gamma & = & \sum_{k} Q_k \left[\left(X^{(1)}_{e_R}\right)^*_{jk} \left(X^{(1)}_{e_R}\right)_{ik} y_{RR} +\left(X^{(1)}_{e_L}\right)^*_{jk} \left(X^{(1)}_{e_L}\right)_{ik} y_{LL} \right. \nonumber \\
 &+& \left. \left(X^{(1)}_{e_R}\right)^*_{jk} \left(X^{(1)}_{e_L}\right)_{ik} y_{RL} +\left(X^{(1)}_{e_L}\right)^*_{jk} \left(X^{(1)}_{e_R}\right)_{ik} y_{LR} \right]\,,\eeqa
and  $c_R^\gamma$ can be obtained with replacement $L \leftrightarrow R$ in the coupling matrices appearing in  the above expression. $Q_k$ denotes electric charge of $l_k$ lepton. The explicit expressions of the loop functions $y_{LL}$, $y_{RR}$, $y_{LR}$ and $y_{RL}$ can be found in \cite{Lavoura:2003xp}. Numbers estimated for ${\rm BR}[\mu \to e \gamma]$, ${\rm BR}[\tau \to \mu \gamma]$ and ${\rm BR}[\tau \to e \gamma]$ using the above expressions are listed in Table \ref{tab:LFV} for three example solutions.

It can be seen from Table \ref{tab:LFV} that the strongest constraints on the flavourful $Z_1$ interactions arise from $\mu$ to $e$ transition and processes like $l_i \to 3 l_j$ as they arise at the tree level. The process $\mu \to e \gamma$ also puts a comparable limit on $M_{Z_1}$. It is seen that $M_{Z_1}$ up to 10 TeV is ruled out by these LFV processes disfavouring the benchmark solution S1. A comparison between the various numbers in Table \ref{tab:meson_WC} and \ref{tab:LFV} indicates that the LFV constraints are less stringent than those arising from meson-antimeson oscillations.

\subsection{Direct and electroweak constraints}
The quark and lepton flavour violations put strong lower bounds on the masses of new particles which more or less supersede the direct search constraints. For example, the latest results from the LHC lead to $M_{Z_1} > 5.15$ TeV for $Z_1$ with ${\cal O}(1)$ flavour diagonal couplings with the SM fermions \cite{CMS:2021ctt}. The limit increases to $M_{Z_1} > 7.20$ TeV if $Z_1$ has generic diquark couplings \cite{CMS:2018mgb}. Similarly, the current direct search constraints on the vectorlike fermions imply $m_B > 1.57$ TeV \cite{CMS:2020ttz,ATLAS:2018mpo} and $m_T > 1.31$ TeV \cite{CMS:2018wpl,ATLAS:2018ziw}. It can be seen from Table \ref{tab:sol} and results of the previous subsections that these constraints are much weaker than the ones imposed by FCNCs.

Another class of constraints arise in the model due to $Z-Z_{1,2}$ mixing as the Higgses are charged under both the SM and extended gauge symmetries. Analogous to \cite{Babu:1997st}, this mixing can be parametrized by mixing angles
\be \label{zzmixing}
\sin \theta_{1,2} = \frac{g_{1,2}}{\sqrt{g^2 + g^{\prime 2}}}\,\left(\frac{M_Z}{M_{Z_{1,2}}}\right)^2\,, \ee
where $g$ and $g^\prime$ are the strengths of $SU(2)_L$ and $U(1)_Y$ gauge interactions, respectively. In the present framework, the fermion mass hierarchy implies $\theta_{2} \ll \theta_{1} \ll 1$ and the dominant effects arise due to $Z-Z_{1}$ mixing. The latter leads to the flavour non-universal couplings to the SM fermions for $Z$ boson also. However, these couplings are suppressed by a factor of $M_Z^2/M_{Z_1}^2$ in comparison to those of $Z_1$.

The $Z-Z_{1}$ mixing modifies the $\rho$ parameter which is precisely measured along with the other electroweak observables. At the leading order in $\theta_1$, the shift in the $\rho$ parameter can be obtained as \cite{Allanach:2021kzj} 
\be \label{delta_rho}
\Delta \rho = \frac{g_1^2}{g^2 + g^{\prime 2}}\,\left(\frac{M_Z}{M_{Z_{1,2}}}\right)^2\,.\ee
A global fit result, $\rho = 1.00039 \pm 0.00019$ \cite{ParticleDataGroup:2020ssz}, then implies $M_{Z_1} \geq 4.5$ TeV for $g_1=1$. Non-zero $Z-Z_{1}$ mixing also modifies the couplings of $Z$ with neutrinos which can be constrained by the invisible decay width of $Z$ boson. This constraint translates to $M_{Z_1}/g_1 \geq 0.95$ TeV \cite{Davighi:2021oel}.  The flavour non-universal couplings of $Z$ to leptons induced by $Z-Z_{1}$ mixing give rise to lepton flavour universality violation in $Z$ decays. The latter is severely constrained by LEP measurements which implies $R = 0.999 \pm 0.003$ \cite{ParticleDataGroup:2020ssz} where $R$ is a ratio of partial decay widths of $Z$ decaying into a pair of electrons and muons. At the leading order in $\theta_1$, the shift in $R$ from unity due to the new physics contributions is given by \cite{Davighi:2021oel}
\be \label{DeltaR}
\Delta R \simeq 4 g_1\,\sin \theta_1\, \frac{g \cos \theta_W - 3 g^\prime \sin \theta_W }{\left(g \cos \theta_W - g^\prime \sin \theta_W \right)^2 + 4 g^{\prime 2}\,\sin^2 \theta_W}\,.\ee
The LEP constraint then leads to $M_{Z_1}/g_1 \geq 1.3$ TeV.

In summary, the constraints from the direct searches and electroweak precision observables are at least two orders of magnitude weaker than those from quark and lepton flavour violating interactions. Various limits discussed in this section suggest a lower bound, $M_{Z_1}/g_1 > 10^3$ TeV, for the generic viable solutions obtained in the present model.

\section{Summary and outlook}
\label{sec:summary}
We explore a mechanism for the radiative induction of the masses for the first and second generation charged fermions. It uses extended abelian gauge symmetry which prevents tree-level masses for all the SM fermions. The third generation fermions can acquire masses with the help of an additional vectorlike family through the seesaw-like mechanism. Subsequently, the radiative corrections induced by spontaneous breaking of extended gauge symmetry can give rise to masses for the remaining fermions explaining their hierarchical spectrum. It is shown that, for the underlying mechanism to work viably, the SM fermions should have flavour non-universal charges under the new symmetry.

Using this general setup, we give an explicit model based on $U(1)_1 \times U(1)_2$ symmetry which is a generalization of well-known $L_\mu - L_\tau$ and $L_e - L_\mu$ symmetries, respectively. The breaking of $U(1)_1$ ($U(1)_2$) induces radiative masses for the second (first) generation fermions and the hierarchy among the masses of the first two families can be attributed to the hierarchy between the breaking scales of two $U(1)$s. We give three example numerical solutions which reproduce the observed charged fermion masses and quark mixing parameters and discuss various constraints from the quark and lepton flavour violations, direct searches and electroweak precision observables on the obtained solutions. Although the radiative mass generation mechanism does not fix unambiguously the scale of new physics, the current constraints imply that the new particles must be heavier than $10^3$ TeV. The requirement of reproducing viable fermion mass spectrum more or less fixes the relative mass scales of new vector bosons and vectorlike fermions.

Although the main motivation for the radiative mass generation mechanism is to make the masses of fundamental fermions calculable parameters of the theory, the results obtained in this paper still involve a large number of free parameters. Unlike in the SM, the fundamental parameters of the model do not span a wide range of magnitude. However, their large number and non-unique values make the model less predictive. One way to improve upon this is to accommodate the $U(1)_1 \times U(1)_2$ symmetry in a larger flavour symmetry based on a single gauge group or to use only one $U(1)$ with appropriate flavour non-universal charges. For the latter, a systematic scan of the fermion spectrum based on anomaly-free charges, similar to the one performed recently in \cite{Patel:2020bwo} in the context of 5D models, would be required. Another interesting possibility is to restrict vertically by unifying various SM quarks and leptons in some irreducible representations of a grand unified theory. Both these approaches can lead to a reduction in the number of free parameters and may provide more predictive models. These alternatives shall be explored in our future works.

\section*{Acknowledgements}
This work is partially supported under MATRICS project (MTR/2021/000049) by the Science \& Engineering Research Board (SERB), Department of Science and Technology (DST), Government of India.

\appendix
\section{Calculation of 1-loop self-energy correction to the fermion masses}
\label{app:loop}
\begin{figure}[!ht]
    \centering
    \includegraphics[width=8cm]{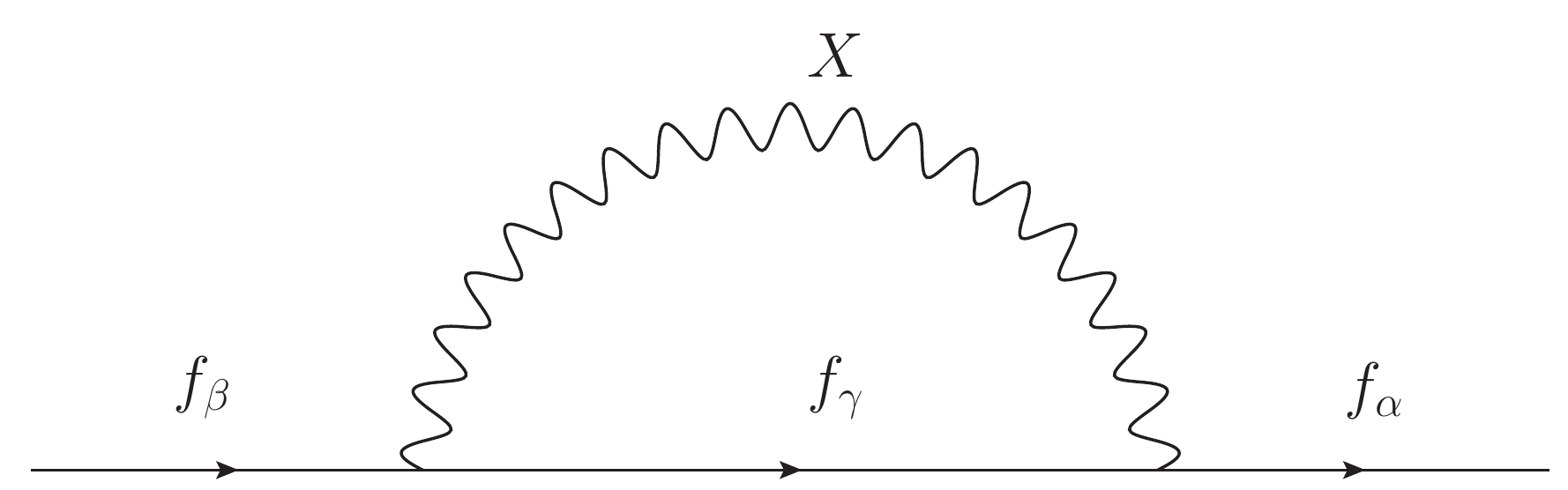}
    \caption{Gauge boson induced fermion self-energy correction at 1-loop}
    \label{fig:oneloop}
\end{figure}

The gauge interactions in Eq. (\ref{L_gauge_2}) can be written as
\beqa \label{a11}
-{\cal L}_{\rm gauge} &=& g_X\, X_\mu  \overline{f}_{ \alpha} \gamma^\mu \,{\cal C}_{\alpha \beta}\,  f_{\beta},\eeqa 
with the coupling defined as
\be \label{a12}
{\cal C}_{\alpha \beta} = ({\cal Q}_L)_{\alpha \beta}\, P_L + ({\cal Q}_R)_{\alpha \beta} P_R \ . \ee
The 1-loop fermion self energy correction induced by the gauge boson in the loop is shown in Fig. \ref{fig:oneloop}.
The amplitude of this diagram is given by
\beqa \label{a13}
-i\Sigma_{\alpha \beta}(p) &=& \sum_{\gamma}\int \frac{d^4 k}{(2\pi)^4}(-ig_X\gamma^\mu {\cal C}^{\dagger}_{\alpha \gamma})\frac{i(\slashed{k}+\slashed{p}+m_{\gamma})}{[(k+p)^2-m_{\gamma}^2 + i\epsilon]}(-ig_X\gamma^\nu {\cal C}_{\gamma \beta})\Delta_{\mu \nu}(k)\,,\eeqa
with 
\be \label{a14}
\Delta_{\mu \nu}(k) = \frac{-i}{k^2-{M_{X}}^2 + i\epsilon} \left( \eta_{\mu \nu}-(1-\zeta)\frac{k_\mu k_\nu}{k^2-\zeta M_X^2}\right)\,.\ee

We set $p=0$ in order to go on the mass shell for the massless fermion and compute the loop contribution in the Feynman-'t Hooft gauge ($\zeta =1$) in dimensional regularization scheme. As the denominator is an even function of $k$, the terms proportional to odd number of $k'$s in numerator vanishes. Therefore,
\begin{align}
\Sigma_{\alpha \beta}(0) =\ -i g_X^2 \mu^\epsilon\ & \sum_{\gamma}\left[({{\cal Q}_L})^\dagger_{\alpha \gamma} P_R + ({{\cal Q}_R})^\dagger_{\alpha \gamma} P_L \right]  \nonumber \\
& \times \int \frac{d^dk}{(2\pi)^d} \frac{d \ m_{\gamma} }{k^2-m_{\gamma}^2 + i\epsilon}\frac{1}{k^2-{M_{X}}^2 + i\epsilon}\ {{\cal C}}_{\gamma \beta} \label{a15}
\end{align} 
where, $d=4-\epsilon$ and we have used the following relations in order to obtain Eq. (\ref{a15}).
\begin{align}
\gamma^\mu {\cal C}^{\dagger}_{\alpha \gamma} &= \left[({{\cal Q}_L})^\dagger_{\alpha \gamma} P_R + ({{\cal Q}_R})^\dagger_{\alpha \gamma} P_L \right] \gamma^\mu\,, \label{a16} \\
 \gamma^\mu \gamma_\mu &= d\,.
\end{align}

Using ${\cal Q}_{L,R}^\dagger = {\cal Q}_{L,R}$ (see Eq. (\ref{calQ})), Eq. (\ref{a15}) can be simplified to 
\begin{align}
\Sigma_{\alpha \beta}(0) =\ \frac{d g_X^2}{16\pi^2}\ & \sum_{\gamma}\ m_\gamma \ \left[({{\cal Q}_L})_{\alpha \gamma} ({{\cal Q}_R})_{\gamma \beta} P_R + ({{\cal Q}_R})_{\alpha \gamma} ({{\cal Q}_L})_{\gamma \beta} P_L \right]  \nonumber \\
& \times \frac{({2 \pi \mu})^\epsilon}{i \pi^2} \int d^dk \frac{ 1 }{k^2-m_{\gamma}^2 + i\epsilon}\frac{1}{k^2-{M_{X}}^2 + i\epsilon}
\end{align}
The integration in the above can be evaluated and expressed in terms of Passarino-Veltmann function $B_0$ \cite{Passarino:1978jh} leading to a final expression
\be \label{a19}
\Sigma_{\alpha \beta}(0) =\  \frac{ g_X^2}{4\pi^2}\  \sum_{\gamma}\ m_\gamma \ \left[({{\cal Q}_L})_{\alpha \gamma} ({{\cal Q}_R})_{\gamma \beta} P_R + ({{\cal Q}_R})_{\alpha \gamma} ({{\cal Q}_L})_{\gamma \beta} P_L \right]\ B_0[M^2_X,m_\gamma ^2]   \ee 
with 
\beqa
B_0[M^2_X,m_\gamma ^2] &=& \frac{({2 \pi \mu})^\epsilon}{i \pi^2} \int d^dk\, \frac{ 1 }{k^2-m_{\gamma}^2 + i\epsilon}\frac{1}{k^2-{M_{X}}^2 + i\epsilon} \, \nonumber \\
& = & \frac{2}{\epsilon}+1-\gamma + \ln{4\pi}-  \frac{M_X^2 \ln M_X^2 - m_\gamma ^2 \ln m_\gamma ^2}{M_X^2 - m_\gamma ^2}\,. \eeqa
The above result is used in Eq. (\ref{Sigma}) to determine explicitly the radiative contributions to the fermion mass matrices.

\section{Scalar potential}
\label{app:scalar}
The most general renormalizable scalar potential of the model, invariant under the SM gauge symmetry and $G_F$, is written as:
\beqa \label{potential}
V &=& m_{u i}^2\, H_{u i}^\dagger H_{u i}\, +\, m_{d i}^2\, H_{d i}^\dagger H_{d i}\, +\, m_{\eta i}^2\, \eta_{i}^\dagger \eta_{i}\, \nonumber \\
& + & \left\{(m_{ud\eta})_{ijk}\, \epsilon_{ijk}\, \eta_i H_{u j} H_{d k}\,+\, (m_{\eta})_{ijk}\, \epsilon_{ijk}\, \eta_i \eta_j \eta_k + {\rm h.c.}\right\}\, \nonumber \\
& + &  (\lambda_u)_{ij}\, H_{u i}^\dagger H_{u i} H_{u j}^\dagger H_{u j}\,+\, (\lambda_d)_{ij}\, H_{d i}^\dagger H_{d i} H_{d j}^\dagger H_{d j}\,+\, (\lambda_\eta)_{ij}\, \eta_{i}^\dagger \eta_{i} \eta_{j}^\dagger \eta_{j}\, \nonumber \\
& + &  (\lambda_{ud})_{ij}\, H_{u i}^\dagger H_{u i} H_{d j}^\dagger H_{d j}\,+\, (\lambda_{u\eta})_{ij}\, H_{u i}^\dagger H_{u i} \eta_j^\dagger \eta_j\,+\, (\lambda_{d \eta})_{ij}\, H_{d i}^\dagger H_{d i} \eta_{j}^\dagger \eta_{j}\, \nonumber \\
& + &  (\tilde{\lambda}_u)_{ij}\, H_{u i}^\dagger H_{u j} H_{u j}^\dagger H_{u i}\,+\, (\tilde{\lambda}_d)_{ij}\, H_{d i}^\dagger H_{d j} H_{d j}^\dagger H_{d i}\, \nonumber \\
& + &  (\tilde{\lambda}_{ud})_{ij}\, H_{u i}^\dagger H_{u j} H_{d j}^\dagger H_{d i}\,+\, (\tilde{\lambda}_{u\eta})_{ij}\, H_{u i}^\dagger H_{u j} \eta_j^\dagger \eta_i\,+\, (\tilde{\lambda}_{d \eta})_{ij}\, H_{d i}^\dagger H_{d j} \eta_{j}^\dagger \eta_{i}\, \nonumber \\
& + & \left\{(\lambda_{ud\eta})_{ij}\, \eta_i^\dagger H_{ui} \eta_j^\dagger H_{dj} + (\tilde{\lambda}_{ud\eta})_{ij}\, \eta_i^\dagger H_{uj} \eta_j^\dagger H_{di} + {\rm h.c.}\right\}\,,\eeqa 
where, $i,j,k = 1,2,3$ are flavour indices. The diagonal elements of all the $\tilde{\lambda}$ matrices can be chosen zero without loss of generality. We assume the general VEVs for various fields, as parametrized in Eqs. (\ref{mu_expl},\ref{mup_expl}), which breaks all symmetries except $U(1)$ corresponding to electromagnetism. Given a large number of parameters in Eq. (\ref{potential}), we assume that such minima exist for a suitable choice of their values.

It can be seen that the potential does not possess any enhanced global symmetry in its most general form. Therefore, one does not find any new Goldstone bosons other than the ones corresponding to the spontaneous breaking of the SM and $G_F$ symmetries which are eaten by the massive $W^\pm$, $Z$ and $Z_{1,2}$ bosons. The potential has an $SU(3)$ global symmetry if all the quadratic, cubic and quartic couplings are assumed flavour universal. This symmetry corresponds to an invariance under a rotation $\Phi_i \to U_{ij} \Phi_{j}$ with $\Phi = H_u, H_d$ and $\eta$. Moreover, for vanishing $\tilde{\lambda}_{ud}$, $\tilde{\lambda}_{u\eta}$, $\tilde{\lambda}_{d \eta}$,  $\lambda_{ud\eta}$, $\tilde{\lambda}_{ud\eta}$, $m_{ud\eta}$ and $m_\eta$, the scalar potential can possess an enhanced $[U(3)]^3$ symmetry corresponding to separate rotations for $H_u$, $H_d$ and $\eta$.

\section{Computed values of flavour violating couplings}
\label{app:Xvalues}
In this Appendix, we give the numerical values of various coupling matrices $X^{(k)}_{f_L}$ and $X^{(k)}_{f_R}$ for benchmark solution 2 (S2). 
\begin{eqnarray}
X^{(1)}_{u_L} &= \left(
\begin{array}{ccc}
 -0.9964 & 0.0597 i & 0 \\
-0.0597 i & -0.0008 & -0.0528 \\
 0 & -0.0528 & 0.9972 \\
\end{array}
\right); 
& X^{(1)}_{u_R} = \left(
\begin{array}{ccc}
 -0.1459 & 0.5747 i & -0.0023 i \\
 -0.5747 i & -0.1466 & -0.7781 \\
 0.0023 i & -0.7781 & 0.2925 \\
\end{array}
\right)\\
X^{(1)}_{d_L} &= \left(
\begin{array}{ccc}
 -0.9262 & 0.2618 i & -0.0034 i \\
 -0.2618 i & -0.0706 & -0.0576 \\
 0.0034 i & -0.0576 & 0.9967 \\
\end{array}
\right); 
 & X^{(1)}_{d_R} = \left(
\begin{array}{ccc}
 -0.2665 & 0.4497 i & 0.0021 i \\
 -0.4497 i & -0.6887 & -0.2625 \\
 -0.0021 i & -0.2625 & 0.9552 \\
\end{array}
\right) \\
X^{(1)}_{e_L} &= \left(
\begin{array}{ccc}
 -0.9978 & 0.0460 & -0.0105 \\
 0.0460 & 0.1330 & 0.3423 \\
 -0.0105 & 0.3423 & 0.8648 \\
\end{array}
\right); 
 & X^{(1)}_{e_R} = \left(
\begin{array}{ccc}
 -0.9741 & 0.1551 & -0.0375 \\
 0.1551 & 0.1169 & 0.3547 \\
 -0.0375 & 0.3547 & 0.8572 \\
\end{array}
\right)
\end{eqnarray}
\begin{eqnarray} 
 X^{(2)}_{u_L} &= \left(
\begin{array}{ccc}
 0.0036 & 0.0594 i & 0.0047 i \\
 -0.0594 i & 0.9909 & 0.1053 \\
 -0.0047 i & 0.1053 & -0.9944 \\
\end{array}
\right) 
 ; X^{(2)}_{u_R}= \left(
\begin{array}{ccc}
 0.5456 & -0.0378 i & 0.6724 i \\
 0.0378 i & -0.3627 & 0.5613 \\
 -0.6724 i & 0.5613 & -0.1829 \\
\end{array}
\right)\\
 X^{(2)}_{d_L} &= \left(
\begin{array}{ccc}
 0.0736 & 0.2603 i & 0.0247 i \\
 -0.2603 i & 0.9200 & 0.1102 \\
 -0.0247 i & 0.1102 & -0.9936 \\
\end{array}
\right); 
 X^{(2)}_{d_R} =\left(
\begin{array}{ccc}
 0.7233 & 0.4203 i & 0.1738 i \\
 -0.4203 i & 0.2262 & 0.2348 \\
 -0.1738 i & 0.2348 & -0.9495 \\
\end{array}
\right) \\
 X^{(2)}_{e_L} &= \left(
\begin{array}{ccc}
 0.0022 & 0.0420 & -0.0205 \\
 0.0420 & 0.7281 & -0.6828 \\
 -0.0205 & -0.6828 & -0.7302 \\
\end{array}
\right);  
X^{(2)}_{e_R}= \left(
\begin{array}{ccc}
 0.0254 & 0.1418 & -0.0704 \\
 0.1418 & 0.6962 & -0.6869 \\
 -0.0704 & -0.6869 & -0.7216 \\
\end{array}
\right)
\end{eqnarray}
It can be seen that both the diagonal and off-diagonal elements are of similar magnitude and no particular pattern or hierarchies in values is seen. We find similar values for the other two solutions.

\bibliography{references}

\end{document}